\documentclass[floatfix,pre,twocolumn,superscriptaddress]{revtex4}
\usepackage{latexsym}
\usepackage{graphicx}
\usepackage{amssymb}
\usepackage{amsmath,enumerate,rotate}

\usepackage{mathptmx}       
\usepackage{helvet}         
\usepackage{courier}        
\usepackage{amsthm,amsmath,amsfonts,amssymb,epsfig,latexsym}

\newcommand{\R}{\mathbb R}

\def\be#1\ee{\begin{equation}#1\end{equation}}
\newcommand{\fer}[1]{(\ref{#1})}

\newcommand{\bq}{\begin{equation}}
\newcommand{\eq}{\end{equation}}

\def\bqa{\begin{eqnarray}}
\def\eqa{\end{eqnarray}}

\newcommand{\bd}{\begin{displaymath}}
\newcommand{\ed}{\end{displaymath}}
\newcommand{\ba}{\begin{eqnarray}}
\newcommand{\ea}{\end{eqnarray}}

\def\R{\mathbb{R}}
\def\var{\varepsilon}

\def\gl{ \hat g}

\begin{document}

\author{F.~Bassetti} \affiliation{Department of Mathematics, University of Pavia,  via Ferrata 1, Pavia, Italy.}
\email[e-mail address: ] {federico.bassetti@unipv.it}
\author{G.~Toscani}
\affiliation{Department of Mathematics, University of Pavia,  via Ferrata 1, Pavia, Italy. } \email[ e-mail address: ]{giuseppe.toscani@unipv.it}

\date{\today}

\title{Explicit equilibria in a kinetic model of gambling}

\begin{abstract}
   We introduce and discuss a nonlinear kinetic equation of Boltzmann type which describes the evolution of wealth in a pure
gambling process, where the entire sum of wealths of two agents is up for gambling, and randomly shared between the agents.  For this equation the
analytical form of the steady states is found for various realizations of the random fraction of the sum which is shared to the agents.  Among
others, Gibbs distribution appears as steady state in case of a uniformly distributed random fraction, while Gamma distribution appears for a
random fraction which is Beta distributed.
The case in which the gambling game is only \emph{conservative-in-the-mean} is shown to lead to an explicit heavy tailed distribution.
\end{abstract}

\maketitle

\section{Introduction}

Various concepts and techniques of statistical mechanics have been fruitfully applied for years to a wide variety of complex extended systems,
physical and otherwise, in an effort to understand the emergent properties appearing in them. Economics is, by far, one of the complex extended
systems to which methods borrowed from statistical mechanics for particle systems have been applied \cite{ Ch02,  ChaCha00, ChChSt05, CoPaTo05,
DY00, Ha02, IKR98, Sl04}. In most of the models introduced so far, the trading mechanism leaves the total mean wealth unchanged. Then, a
substantial difference on the final behavior of the model (presence or not of tailed steady states) can be observed depending on the fact that
binary trades are pointwise conservative, or conservative in the mean \cite{DMT, MaTo07}. The asymptotic distribution of wealth, however, depends
completely on the microscopic structure of binary trades. Other kinetic models have been recently proposed, which, while maintaining the kinetic
description, introduce more sophisticated rules for trading. For example, a description of the behavior of a stock price has been developed by
Cordier, Pareschi and Piatecki in \cite{CPP}. Further, there have been efforts to include non-microscopic effects, like global taxation (and
subsequent redistribution), in recent works of Guala \cite{Gua}, Pianegonda, Iglesias, Abramson and Vega \cite{Pia}, Garibaldi, Scalas and
Viarengo \cite{GSV} and Bisi, Spiga and the present author \cite{BST, To2}.

Despite the high number of studies devoted to the subject, well documented by various recent review papers \cite{CC09, gup,Pat, yak, YR},
analytical solutions or explicit steady states for wealth distribution densities are rarely present in the literature. The few exceptions are the
self-similar solution with Pareto tails found by Slanina \cite{Sl04} for a kinetic model of a non-conservative (decreasing in the mean) economy,
and the steady state solution (Gibbs distribution) of a linear kinetic equation modelling taxation and uniform redistribution \cite{To2}.

The goal of this paper is to show that in a pure gambling trade market explicit equilibrium solutions can be obtained by suitably choosing the
random fraction of money which governs the microscopic interaction.

Wealth exchange processes in a discrete market are characterized by binary trades. A fixed number of $N$ agents in a system are allowed to interact (trade)
stochastically and thus wealth is exchanged between them. The basic steps of such a wealth exchange model can be described as follows
 \be\label{coll0}
 w_i(t + 1) = w_i(t) + \Delta w, \qquad  w_j(t + 1) = w_j(t)- \Delta w,
 \ee
 where $w_i(t)$ and $w_j (t)$ are wealths of
$i$-th and $j$-th agents at time $t$ and $w_i(t+1)$ and $w_j (t+1)$ are that at the next time step $t+1$. The amount $\Delta w$ (to be won or to
be lost by an agent) is determined by the nature of interaction. If the agents are allowed to interact for a long enough time, a steady state
equilibrium distribution for individual wealth is achieved. The equilibrium distribution does not depend on the initial configuration (initial
distribution of wealth among the agents).

In a pure gambling process \cite{DY00}, the entire sum of wealths of two agents is up for gambling. Some random fraction of this sum is shared by
one agent and the rest goes to the other. The randomness is introduced into the model through a parameter $\var$ which is a random number drawn
from a probability distribution in $[0, 1]$. In general it is assumed that $\var$ is independent of a pair of agents, so that a pair of agents is
not likely to share the same fraction of aggregate wealth the same way when they interact repeatedly. The interaction rule can be seen through
 \be\label{coll}
 \begin{split}
 &w_i(t + 1) = \var [w_i(t) + w_j (t)], \\ &   w_j(t + 1) = (1 - \var)[w_i(t) + w_j(t)],\\
\end{split}
 \ee
 where the pair of agents (indicated by i and j)
are chosen randomly. The amount of wealth that is exchanged is now
$\Delta w = \var [w_i(t) + w_j (t)] - w_i(t)$. Numerical experiments
\cite{gup} show that, if $\var$ is a random number drawn from a
uniform distribution in $[0, 1]$, the individual wealth distribution
at equilibrium emerges out to be Gibbs distribution like
exponential. It is strongly believed, and somewhat argued from
numerical experiments, that a slight variant of the previous model
produces at equilibrium Gamma-like distributions \cite{CP08}, but,
as remarked by the authors, the form of the exact solution is still
an open question. The main feature  of trade \fer{coll} is such the
amount of money which is restituted in a single trade coincides with
the amount of money the two agents use for gambling (\emph{pointwise
conservative trade}).  In this gambling market there is no rule of
the (conserved) total amount of money initially in the hands of the
agents. In other words, agents do not take advantage from the amount
of money available in the market.

The idea of using this money as a \emph{reservoir} for trades can be
easily implemented allowing agents to trade with random profit
 \be\label{coll-r}
  \begin{split}
 & w_i(t + 1) = \var_1 [w_i(t) + w_j (t)], \\ &  w_j(t + 1) = \var_2[w_i(t) + w_j(t)], \\
 \end{split}
 \ee
where the parameters $(\var_1, \var_2) $ are now random numbers drawn from a joint probability distribution  
 such that
 \be\label{cons}
  \langle \var_1 + \var_2 \rangle =1,
\ee where $\langle \cdot \rangle$ denotes as usual the mathematical
expectation. Within this picture, $w_i(t + 1) + w_j(t + 1)$ can be
strictly less than $w_i(t) + w_j (t)$, and in this case the lost
money is achieved by the market, or the reverse situation is
verified, and the additional money is taken from the market.
Condition \fer{cons}, however, guarantees that in the mean the
wealth present in the market is left unchanged. Note that if one
assumes that $\var_i\geq \gamma>0$, $i=1,2$ then
%
agents are prevented from  loosing all their money in a single
trade. Note also that, choosing $(\var_1,\var_2)=(\var,1-\var)$ the
collision rule \eqref{coll-r} becomes \eqref{coll}.

In order to produce a fair game, it will be assumed that the random
numbers $\var_1$ and $\var_2$ are identically distributed. For
trades of type \fer{coll}, this assumption simply forces the random
fraction $\var$ to be symmetric with respect to the value $1/2$.

At a continuous level, the evolution of the wealth distribution can
be described by means of a nonlinear kinetic model of Boltzmann
type, in which the wealth distribution of the market is driven by
\emph{collisions} of type \fer{coll} \cite{Cer94}. Since the total
mean wealth is maintained constant in time, the wealth distribution
is known to converge exponentially towards a steady profile
\cite{DMT, MaTo07}, which depends on the details of the trade
mechanism through $\var_i$. It is well-known that the continuous
description takes advantage from the  the possibility to make use of
the Laplace transform version of the kinetic equation \cite{Bob88}.

Owing to this representation, we will prove that various cases are
explicitly solvable, and lead to an analytic expression of the
steady profile. In particular, choosing $\var_i$ to be  symmetric
Beta distributions one obtains as equilibrium a Gamma distribution
with a parameter which depends on the parameter of the Beta
distribution. A special case emerges here, since Gibbs distribution
emerges for $\var_i$ uniformly distributed. Also, the case of the
\emph{winner takes all} game can be studied in details as limit of
Beta $\var$, with parameter tending to zero. In all these cases,
however, the equilibria possess moments of any order.

Interestingly enough, the case of the \emph{conservative in the
mean} gambling trade \fer{coll-r} can be treated likewise, by
choosing the random variables $\var_i$, $i=1,2$ to be inverse Beta.
In this case, however, the steady state distribution is shown to be
a generalized Gamma distribution with fat tails, which contains as a
particular case the distribution found by Slanina \cite{Sl04} in a
different context. This founding clarifies through a simple example
the role of the \emph{social} use of the money present in the
market. The possibility to access to the money available in the
common \emph{reservoir} allows the formation of the rich class.

The paper is organized as follows. In the next section we introduce
the continuous model which is described by a nonlinear kinetic
equation of Boltzmann type, and its main features are discussed in
some detail. The analytical solutions in the case of the pure
gambling trade \fer{coll} are described in Section 3. Section 4
deals with the gambling rule \fer{coll-r}.

\section{A continuous kinetic model}\label{model}
\setcounter{equation}{0}

Given a fixed number of $N$ agents in a system, which are allowed to
trade, the interaction rules \fer{coll-r} describe a stochastic
process of the vector variable $(v_1(t), \dots, v_N(t))$ in discrete
time $t$. Processes of this type are thoroughly studied e. g. in the
context of kinetic theory of ideal gases. Indeed, if the variables
$v_i$ are interpreted as energies corresponding to $i$-th particle,
one can map the process to the mean-field limit of the Maxwell model
of elastic particles \cite{Cer, Cer94}. The full information about
the process in time $t$ is contained in the $N$-particle joint
probability distribution $P_N(t, v_1, v_2, \dots, v_N)$. However,
one can write a kinetic equation for one-marginal distribution
function
 \[
P_1(t,v) = \int P_N(t, v, v_2, \dots, v_N)dv_2\cdots dv_N ,
 \]
 involving only one- and two-particle distribution functions
\cite{Cer, Cer94}.
 \[
 \begin{split}
  P_1(t+1,v)- &P_1(t,v) =
\Big \langle \frac 1N \Big[ \int P_2(t, v_i, v_j)\Big( \delta( v - \var_1(v_i + v_j))
\\ & + \delta( v - \var_2)(v_i + v_j)) \Big) dv_i \, dv_j -
2P_1(t, v)\Big]\Big \rangle , \\
 \end{split}
 \]
which may be continued to give eventually an infinite hierarchy of
equations of BBGKY type \cite{Cer}.  The standard approximation,
which neglects the correlations between the wealth of the agents
induced by the trade gives the factorization
 \[
P_2(t, v_i, v_j) =P_1(t,v_i)P_1(t,v_j),
 \]
which implies a closure of the hierarchy at the lowest level. In
fact, this approximation becomes exact for $N \to \infty$.
Therefore, in thermodynamic limit the one-particle distribution
function bears all information. Rescaling the time as $\tau = 2t/N$
in the thermodynamic limit $N \to \infty$, one obtains for the
one-particle distribution function $f(v,t) = P_1(v,t)$ the
Boltzmann-like kinetic equation
 \be\label{Boltz}
 \begin{split}
\frac{\partial f (t, v)}{\partial t} & = \frac 12 \Big \langle \int f(t, v_i)f(t, v_j)
 \Big ( \delta( v - \var_1(v_i + v_j)) \\ & + \delta( v - \var_2)(v_i
+ v_j)) \Big ) dv_i \, dv_j\Big \rangle - f(t,v), \\
 \end{split}
 \ee
which describes the process \fer{coll-r} in the limit $N \to \infty$.

Owing to equations of type \fer{Boltz}, the study of the
time-evolution of the wealth distribution among individuals in a
simple economy, together with a reasonable explanation of the
formation of tails in this distribution has been recently achieved
in \cite{MaTo07} (see also \cite{DMT, DMT1}). The Boltzmann-like
equation \fer{Boltz} can be fruitfully written in weak form. It
corresponds to say that the solution $f(v,t)$ satisfies, for all
smooth functions $\phi(v)$
 \be
  \label{kine-w}
 \begin{split}
& \frac{d}{dt}\int_{\R_+}f(t,v)\phi(v)\,dv  =  \frac 12 \Big \langle \int_{\R_+^2}
\bigl( \phi(v^*)+ \phi(w^*)\\ & \quad -\phi(v) - \phi(w) \bigr) f(t,v)f(t,w)
\,dv \, dw \Big \rangle  , \\
 \end{split}
 \ee
where the post-trade wealths $(v^*, w^*)$ obey to the rule \fer{coll-r}
 \be\label{colli}
 v^* = \var_1 (v+w), \qquad  w^* = \var_2(v+w).
 \ee
Note that \fer{kine-w} implies  that $f(v,t)$ remains a probability density if it so initially
 \be\label{m0}
\int_{\R_+} f(t, v)\,dv = \int_{\R_+} f_0(v)\,dv = 1 .
 \ee
Moreover, on the basis of \fer{cons}, the choice $\phi(v) = v$ shows
that also the total mean wealth is preserved in time
 \be\label{mm}
 m(t) = \int_{\R_+}v f(t,v)\,dv = \int_{\R_+}v f_0(v)\,dv = m(0) .
 \ee
Consequently, without loss of generality, in what follows, we assign to the initial density a unit mean
 \be\label{nor}
\int_{\R_+} v f_0(v)\,dv = 1 .
 \ee
Setting   $\phi(v) = \exp\{-\xi v\}$ in \eqref{kine-w},
\cite{Bob88}, one gets the Boltzmann equation for the Laplace
transform $\hat f$ of $f$, where
 \[
\hat f(t,\xi)=\int_{\R^+} e^{-\xi v}f(t,v)dv.
 \]
Direct computations \cite{MaTo07} show that $\hat f(t,\xi)$
satisfies the equation
\begin{equation}
\frac{\partial \hat f(t,\xi)}{\partial t} +\hat f(t,\xi) = \frac{1}{2}
 \left\langle   \hat f(t,\xi \var_1)^2+\hat f(t,\xi\var_2)^2 \right\rangle .
\end{equation}
As extensively discussed in \cite{DMT, MaTo07},  explicitly computable conditions on the gambling variables $\var_i$, $i =1,2$,  guarantee that
the distribution $f(t,v)dv$ converges weakly to a \emph{universal} probability distribution whose Laplace transform  is the unique solution of
\begin{equation}\label{stazeq}
\hat f_\infty(\xi) = \frac{1}{2}\left\langle
 \hat f_\infty(\xi \var_1)^2+\hat f_\infty(\xi\var_2)^2
 \right\rangle
\end{equation}
with
\[
\hat f_\infty^{\,\,'}(\xi)|_{\xi=0}=-1.
\]
This fact follows from Thm. 3.3  in \cite{MaTo07} (see also Thm. 2 in \cite{BLM10}), which links both the convergence and the boundedness of
moments of the equilibrium solution to the sign of the key function $G(s)$, defined as
\[
G(s):=\langle \var_1^s+\var_2^s\rangle -1
\]
In particular the convergence result is valid without additional assumption on $f_0$
provided that  $G \,\,'(1)<0$.

By further requiring that the random variables $\var_1$ and $\var_2$ are distributed with the same law, the equilibrium solution $\hat f_\infty$
is found to be the unique solution to
\begin{equation}\label{staz-eq}
\hat f_\infty(\xi) =\left\langle
 \hat f_\infty(\xi \var)^2
 \right\rangle,
\end{equation}
where the random variable $\var$ is distributed according to the common law of $\var_1$ and $\var_2$. Equation \fer{staz-eq} can be better
understood by saying that, if $Z$ is a random variable with law $f_\infty$, the law of $Z$, defined in \fer{staz-eq}, is a distributional fixed
point of the equation
 \be\label{per1}
 Z =^d \var(Z_1 +Z_2),
 \ee
where $=^d$ means identity in distribution and one assumes that the random variables $Z_1,Z_2$ and $Z$ have
the same probability law, while the variables $Z_1,Z_2$ and $\var$ are
assumed to be stochastically independent. Equations of type \fer{per1}
are well-known and extensively studied, see e.g. \cite{Ligg, Liu}.

\section{The pure gambling trade model}\label{pure}

Let us start from the pure gambling case in which
$(\var_1,\var_2)=(\var,1-\var)$. Having in mind that numerical
experiments are usually done with a random number drawn from a
uniform distribution in $[0, 1]$, which leads the discrete market to
a Gibbs distribution at equilibrium \cite{gup},  we fix the random
number $\epsilon$ to be a symmetric Beta random variable of
parameters $(a,a)$, with $(a>0)$. We recall that a random variable
is Beta distributed with parameters $(a,b)$ if its density is
\[
 \beta_{a,b}(x)=\frac{\Gamma(a+b)}{\Gamma(a)\Gamma(b)} x^{a-1}(1-x)^{b-1} \qquad x \in (0,1),
\]
see, for instance,  \cite{fristedgray}.

The case $a=1$, where $\epsilon$ is a random number uniformly
distributed on $(0,1)$, confirms the numerical outcome. In this
case, in fact,  \eqref{staz-eq} becomes
 \[
 \hat f_\infty(\xi) = \int_0^1 \hat f^2_\infty(\xi x )\, dx .
 \]
It is easy to see that
\[
\hat f_\infty(\xi)=(1+\xi)^{-1}
\]
is a solution of \eqref{staz-eq}, such that $\hat f_\infty^{\,\,'}(\xi)_{|\xi=0}=-1$. Since $(1+\xi)^{-1}$ is the Laplace transform of the Gibbs
distribution of unit mean
\[
f_\infty(v)=e^{-v}  \qquad (v\geq 0),
\]
Gibbs distribution results as analytical steady solution to the pure gambling trade market, in case $\var$ is a uniform random number in $(0,1)$.

To treat the more general case it suffices to recall that, if $\alpha_1,\alpha_2,\xi>0$, then formula 3.197.4 in \cite{gradshteyn} gives
 \be\label{integrale}
  \begin{split}
  \frac{\Gamma(\alpha_1+\alpha_2)}{\Gamma(\alpha_1)\Gamma(\alpha_2)}
  & \int_{0}^1 \frac{ x^{\alpha_1-1}
  (1-x)^{\alpha_2-1}}{(1+\xi x )^{\alpha_1+\alpha_2}}dx\\ &=(1+ \xi)^{-\alpha_1}.\\
 \end{split}
\ee
Using  identity \fer{integrale} with $\alpha_1=\alpha_2 =a$, one
obtains that  for a general $a>0$
 the function
 \be\label{gam}
 \hat f_\infty(\xi)= \Big (1+\xi \frac{1}{a} \Big )^{-a}
  \ee
solves equation \fer{staz-eq}.  Hence, the equilibrium solution
results in a Gamma distribution of unit mean, with shape parameter
$a$ and scale parameter $1/a$
 \be\label{gam1}
f_\infty(v)=\frac{a^av^{a-1}e^{-av}}{\Gamma(a)} \qquad (v >0).
 \ee
To verify that the results of \cite{MaTo07} and \cite{BLM10}, which guarantee the convergence to the steady state together with its moment
boundedness properties hold, one needs to show that $G \,\,'(1)<0$ holds true  for
\[
G(s):=2\langle \var^s \rangle -1,
\]
where $\var$ is a symmetric Beta random variable of parameters
$(a,a)$, with $(a>0)$. In this case, however,
 \[
G \,\,'(1) = \left\langle \var \log \var \right\rangle <0
 \]
is a simple consequence of the fact that $0 < \var < 1$ with probability one.

 The uniform distribution ($a=1$) appears like
a natural separation between two different behaviors of the equilibrium solutions. In case $a<1$, Gamma distributions \fer{gam1} are monotonically
decreasing, starting from $f_\infty(0)= +\infty$. In the opposite case $a>1$, there is appearance of a peak \emph{around} the unit mean value. The
average wealth is unchanged for every $a$ but, when $a>1$, the number of agents with a wealth closer to the average value increases or, in other
words, the wealth distribution becomes more fair for larger $a$. Analogously, for $a<1$ the distribution gives more weight to agents with a wealth
close to zero.

As a consequence, measures of the inequality of the wealth distribution, such as the Gini coefficient, increase for decreasing $a$ and tend to
zero for $a \to +\infty$. Some insight can be gained by the simple computation of the variance of $f_\infty$. It holds
\[
Var(f_\infty)=\int_{\R^+}v^2f_\infty(v) dv -1= \frac{1}{a}.
\]
Hence the variance (the spreading) decreases as $a$ increases.

There are two interesting limiting cases. The first one is obtained by letting $a \to +\infty$. In this case  $\epsilon$ converges in distribution
to the constant value $1/2$ and one can immediately see that the steady state is a degenerate distribution concentrated on the value $1$ of the
mean wealth. This corresponds to a perfectly fair distribution, in which all agents end up with the same wealth. The other limit case is the
\emph{winner takes all} game, which can be obtained letting $a \to 0$.  In this case the steady state is concentrating on $0$, while its variance
is blowing up. This corresponds to the discrete situation in which a finite number of agents end up with no money, except one which \emph{takes
all}. In the continuous case, the value $a=0$ can not be assumed directly, since the exchange of the limits $a \to 0$ and $t \to \infty$ is not
allowed by the lack of regularity of the equilibrium solution for $a =0$.

\section{Analytic equilibria with heavy tails}\label{model2}

Let us now consider the gambling rule \fer{coll-r}. In agreement with the previous section, we assume that
 \be\label{ga}
\var_i=\frac{1}{4\theta_i}, \qquad i=1,2,
 \ee
where, for a given $a>1$,  $\theta_i$ $i=1,2$ is a Beta random variable of parameters $(a+1/2,a-1/2)$. It is immediate to reckon  that the random
variables $\var_i$ are such that $<\var_i>=1/2$ $i=1,2$. Consequently \eqref{mm} holds and the model is conservative in the mean.  We can invoke
again Thm. 3.3  in \cite{MaTo07} to prove that there is a unique equilibrium solution such that its Laplace transform satisfies
\begin{equation}\label{stazeq2}
\hat f_\infty(\xi) =
\int_{0}^1 \hat f_\infty(\xi/(4x))^2
 \beta_{a+1/2,a-1/2}(x) dx ,
\end{equation}
with \( \hat f_\infty^{\,\,'}(\xi)|_{\xi=0}=-1 \). In this case to apply the results of \cite{MaTo07} and \cite{BLM10}, which guarantee the
convergence to the steady state together with its moment boundedness properties, one needs to show that $G \,\,'(1)<0$ holds true  for
\be\label{g-mod2}
\begin{split}
G(s)&:= \langle \epsilon_1^s+\epsilon_2^s \rangle-1 \\
&= 2\int_0^1 \frac{1}{(4x)^s} \beta_{a+1/2,a-1/2}(x)dx-1 \\
&=\frac{2^{1-2s}\Gamma(2a) \Gamma(a-s+\frac{1}{2})}{\Gamma(2a-s)\Gamma(a+\frac{1}{2})}-1.\\
\end{split}
\ee
The proof of this condition is not direct. For the sake of brevity, we postpone the computations to the Appendix.

Now, we shall prove that the solution of \eqref{stazeq2} is obtained by taking the Laplace transform of the so called Inverse-Gamma distribution \cite{steutel}
of shape parameter $a$ and scale parameter $a-1$, that is
 \be\label{invGamma}
 f_\infty(v)=\frac{(a-1)^a}{\Gamma(a)}
\frac{e^{-\frac{(a-1)}{v}}}{v^{a+1}},
 \ee
 which is peaked around the mean value $1$ and has heavy tails, in that it decays at infinity like $v^{-(a+1)}$.
An analytical solution of type \fer{invGamma}, with a polynomial decay  corresponding to $a = 3/2$ has been discovered by Slanina \cite{Sl04} as
self-similar solution of a kinetic model of a non-conservative (decreasing in the mean) economy.  Motivated by the analogy with a dissipative
Maxwell gas, in \cite{Sl04}  a decreasing in the mean wealth model  where
 \be\label{coll3}
 \begin{split}
& v^* = pv+qw, \quad w^* = qv +pw ; \\ &  p\ge q >0, \quad \sqrt p + \sqrt q = 1 \\
\end{split}
 \ee
has been discussed. Note that, within condition \fer{coll3} on the mixing parameters $p$ and $q$,
 \[
 v^* + w^* = \left(1- 2\sqrt{pq}\right)(v+w) < v+w ,
 \]
which implies that the mean value $m(t)$ at time $t$ decays exponentially to zero at the rate $2\sqrt{pq}$. The standard way to look for
self--similarity is to scale the solution. More precisely, define the rescaled solution $g$ by \be g(t,v) =m(t)f\big(t,m(t)v\big), \ee which
implies that $\int v g(t,v)\,dv= 1$ for all $t \geq 0$.

In terms of the Laplace transform $\gl$ of $g$, it is found that the equation satisfied by $\gl$ reads \cite{DMT1}
 \be \label{laplace}
 \frac{\partial \gl}{\partial t} + \xi (p+q-1)\frac{\partial \gl}{\partial \xi}
=\gl(p\xi)\gl(q\xi ) - \gl(\xi).
 \ee
 Steady solutions to equation \fer{laplace} satisfy
 \be \label{s.laplace} \xi (p+q-1)\frac{\partial \gl}{\partial
\xi }=\gl(p\xi)\gl(q\xi) - \gl(\xi).
 \ee
Direct computations then show that the function
 \be\label{stead1}
 \gl_\infty(\xi) = \left( 1 + \sqrt{2\xi} \right) e^{-\sqrt{2\xi}}
 \ee
solves \fer{s.laplace} for all values of $p$ and $q$ satisfying the constraint $\sqrt p + \sqrt q = 1$. Note that \fer{stead1} is the explicit
Laplace transform of
 \be\label{slan}
 g_\infty(v)=\frac{(1/2)^{3/2}}{\Gamma(3/2)}
\frac{e^{-\frac{1}{2v}}}{v^{5/2}}.
 \ee
Let us set $p =q = 1/4$ in \fer{s.laplace}. Then the steady solution \fer{stead1} satisfies
 \be \label{1.laplace} -\frac{\xi}2 \frac{\partial \gl}{\partial
\xi } + \gl(\xi) = \gl\left(\frac{\xi}4\right)^2 .
 \ee
Following \cite{BCG}, Sect. $6$,  \fer{1.laplace} can be equivalently written in integral form as
 \be\label{int1}
 \gl(\xi) =\int_0^1  \gl\left(\frac{\xi}4\rho^{-1/2}\right)^2\, d\rho,
 \ee
or, setting $\rho^{1/2}= x$
 \be\label{int2}
 \gl(\xi) =\int_0^1 2 \gl\left(\frac{\xi}{4x}\right)^2 x\, dx,
 \ee
which is nothing but \fer{stazeq2} with $a = 3/2$. Consequently the distribution \fer{slan} solves \fer{stazeq2} with $a=3/2$. This argument
establishes a connection between the present problem and the non-conservative one introduced by Slanina \cite{Sl04}, which leads to explicit
computations.

In order to prove that the Laplace transform of \eqref{invGamma} is the solution of \eqref{stazeq2},  since a direct prove seems not
straightforward like in the previous case, we recast the problem in a more probabilistic way.
First of all, let us note that an Inverse--Gamma random variable $Y$ of parameter $(a,a-1)$
can be obtained by taking  $Y= 1/X$ where $X$ has
Gamma distribution of parameter $(a,1/(a-1))$. Recall that  $X$ is a random variable $Gamma(a,1/(a-1))$
if its density is
$$\frac{(a-1)^a  v^{a-1} e^{(a-1)v}}{\Gamma(a)}.$$
Recall also that its
 Laplace transform reads
 \be\label{gam2}
 (1+\xi /(a-1))^{-a}.
 \ee
As discussed in Section \ref{model}, equation \eqref{stazeq2} can be rewritten in equivalent way as
 \be\label{stazeq3}
 Y=^d\frac{1}{4\theta} [Y_1+Y_2]
  \ee
  where $Y_1,Y_2,\theta$ are
independent random variables, $Y,Y_1,Y_2$ have density $f_\infty$,  while $\theta$ has density $\beta_{a+1/2,a-1/2}$. The script $=^d$ has to be
meant as an identity in distribution. To prove \eqref{stazeq3} it suffices to show that $Y^{-1}=^d4\theta[Y_1+Y_2]^{-1}$ or, equivalently,
 \be\label{stazeq4}
 X=^d\frac{4\theta X_1X_2 }{X_1+X_2}
  \ee
where $X,X_1,X_2$ are independent $Gamma(a,1/(a-1))$ random variables. The result follows if one is able to show that $(X_1+X_2)(4\theta
X_1X_2)^{-1}$ has Laplace transform \eqref{gam2}. Setting $G:=X_1+X_2$ and $B:=X_1/(X_1+X_2)$ one rewrites \fer{stazeq4} as
 \be\label{stazeq5}
X=^{d}4 \theta G   B(1-B).
 \ee
It is a classical result of probability theory that, given $X_1$ and $X_2$ which are independent and Gamma distributed, the random variables $G$
and $B$ are stochastically independent and, moreover, that $G$ has $Gamma(2a,1/(a-1))$ distribution, while $B$ has $Beta(a,a)$ distribution. For
the proof of this property, we refer for instance to Chapter 10.4 in \cite{fristedgray}. Moreover, using relation \eqref{integrale} one can reckon
that $\theta G$ has $Gamma(a+1/2,1/(a-1))$ distribution simply by computing its Laplace transform. Using now the fact that $\theta G$ and $B$ are
mutually independent, one can write the Laplace transform of $4 \theta G   B(1-B)$ in the point $\xi$ as
\[
\begin{split}
L(\xi)& =\int_0^1 \frac{\beta_{a,a}(x)}{\big[1+4(a-1)^{-1}\xi x(1-x)\big]^{a+\frac{1}{2}}}\, dx \\
& =2\int_0^{\frac{1}{2}}
\frac{\beta_{a,a}(x)}{\big[1+4(a-1)^{-1}\xi x(1-x)\big]^{a+\frac{1}{2}}} \, dx. \\
\end{split}
\]
At this stage, a simple change of variable shows that
\[
\begin{split}
L(\xi) & =\frac{\Gamma(2a)}{2^{2a-1}\Gamma(a)^2} \int_0^1 \frac{z^{a-1}(1-z)^{\frac{1}{2}-1}}{(1+\xi z/(a-1)))^{a+\frac{1}{2}}}\, dz  \\
&=\int_0^1
\frac{1}{(1+ \xi z/(a-1)))^{a+\frac{1}{2}}} \beta_{a,1/2}(z)\, dz. \\
\end{split}
\]
The last identity follows by the duplication formula
 $\Gamma(2a)=2^{2a-1}\Gamma(a)\Gamma(a+1/2)/\Gamma(1/2)$  (see 8.335.1 in \cite{gradshteyn}).
 Using relation \eqref{integrale} once again we get
$L(\xi)=(1+\xi /(a-1))^{-a}$ and \eqref{stazeq4} is proved.

Some remarks are in order. Within the choice \fer{ga}, the conservative in the mean trade \fer{coll-r} is such that the two agents maintain at
least $1/4$ of the total wealth used  to trade. Thus, trade \fer{coll-r} is in a sense less risky than the conservative trade \fer{coll}, where
one of the two agents can exit from the trade with almost no money. In addition, it follows that the number of moments of the explicit equilibrium
state which are finite increase with $a$. On the other hand, when $a$ increases,  the area described  by the distribution of the random fraction
$\var$ on the interval $[1 , + \infty)$ decreases, and the probability to use wealth of the society is also decreasing. Hence a fat Pareto tail is
obtained through a big use of the common wealth.

\section{Conclusions}

In this paper, we introduced and discussed the equilibrium solution of a nonlinear kinetic equation of Boltzmann type, modelling redistribution of
wealth in  a simple market economy in which trades are described by a standard gambling game. Due to the simplicity of the game trade, analytical
solutions can be obtained in the case in which the post-trade wealths depend on the pre-trade ones through random variables which are Beta
distributed. Previously known analytical solutions are here shown to exit for particular values of the underlying parameters. Despite its
simplicity, the model enlightens the role of the interaction in producing Pareto tails.

\appendix
\section{Appendix}
Let us prove that $G\,'(1)<0$ when $G$ is defined as in \eqref{g-mod2}.
Starting from \eqref{g-mod2}, differentiation shows that
\[
\begin{split}
G{\,'}(s)&=-2^{1-2s}\frac{\Gamma(2a) \Gamma(a-s+\frac{1}{2})}{\Gamma(2a-s)\Gamma(a+\frac{1}{2})} \cdot
\\
&\cdot \Big\{ 2\log(2)+\psi(a- s+\frac{1}{2})-\psi(2a-s) \Big\} \\
\end{split}
\]
where  $\psi(x)=\Gamma{\,'}(x)/\Gamma(x)$ is the Digamma function \cite{gradshteyn}. Using the duplication formula
$2\psi(2x)=\psi(x)+\psi(x+1/2)+2\log(2)$, see 8.365.6 \cite{gradshteyn}, one obtains
\[
G{\,'}(1)=-\frac{1}{4}\frac{\Gamma(2a) \Gamma(a-s+\frac{1}{2})}{\Gamma(2a-s)\Gamma(a+\frac{1}{2})}
Q(a)
\]
where $Q(a):=2\log(2)+\psi(a-1/2)+\psi(a)$.

Now $\psi(1/2)=-\gamma-2\log2$ and $\psi(1)=-\gamma$ ($\gamma$ being the Eulero-Mascheroni constant), see 8.366.1/2 \cite{gradshteyn}, and then
$Q(1)=0$.

The classical expansion formula (see 8.363.8 \cite{gradshteyn})
$$\psi'(x)=\sum_{k \geq 0}\frac{1}{(x+k)^{2}}$$
allows to conclude that, for every $a>1$
\[
Q{\,'}(a)=\sum_{k \geq 0} \frac{1}{(a-1/2+k)^{2}}-\sum_{k \geq 0} \frac{1}{(a+k)^{2}}>0.
\]
 Hence $Q(a)$ is strictly monotone  in $[1,+\infty)$ and since $Q(1)=0$ it follows that
 $Q(a)>0$ for every $a>1$. This shows that $G\,'(1)<0$.

\medskip
{\bf Acknowledgement.} This work has been done under the activities of the National Group of Mathematical Physics (GNFM). The support of the MIUR
projects ``Bayesian methods: theoretical developments and novel applications'' and ``Variational, functional-analytic, and optimal transport
methods for dissipative evolutions and stability problems'' is kindly acknowledged.


\medskip





\begin{thebibliography}{10}

\bibitem{BLM10} F. Bassetti, L. Ladelli, D. Matthes.
Central limit theorem for a class of one-dimensional kinetic equations
\emph{Prob. Theor. Rel. Fields} DOI:10.1007/s00440-010-0269-8
(2010)

\bibitem{BST}
M. Bisi, G. Spiga, G. Toscani. Kinetic models of conservative economies with wealth redistribution, \emph{Commun. Math. Sci.} \textbf{7} (4)
(2009) 901ï¿½-916

\bibitem{Bob88}
  A.V. Bobylev.
  The theory of the spatially Uniform Boltzmann equation for Maxwell molecules.
  {\em Sov. Sci. Review C} \textbf{7},  112--229 (1988).


\bibitem{BCG}
A.V. Bobylev, C. Cercignani, I.M. Gamba. On the Self-Similar Asymptotics for Generalized Nonlinear Kinetic Maxwell Models. \emph{Commun. Math.
Phys.} \textbf{291} 599–-644 (2009).


\bibitem{Cer}
C. Cercignani. The Boltzmann equation and its applications,
\newblock {\em Springer Series in Applied Mathematical Sciences},
  Vol.\textbf{67} Springer--Verlag 1988.

\bibitem{Cer94}
C. Cercignani, R. Illner, M. Pulvirenti,
\newblock The mathematical theory of dilute gases,
\newblock {\em Springer Series in Applied Mathematical Sciences},
  Vol.\textbf{  106} Springer--Verlag 1994.


\bibitem{Ch02}
A.~Chakraborti. {Distributions of money in models of market economy}, {\em Int.
J. Modern Phys. C} \textbf{13}, 1315--1321 (2002).



\bibitem{ChaCha00} A.~Chakraborti, B.K.~Chakrabarti. Statistical
  Mechanics of Money: Effects of Saving Propensity, {\em Eur. Phys. J. B}
  \textbf{17}, 167-170 (2000).

\bibitem{CC09}
A.S. Chakrabarti, B. K. Chakrabarti. Statistical Theories of Income and Wealth Distribution. \emph{Economics, Open-accessment E-Journal}
http://www.economics-ejournal.org/economics/journalarticles/2010-4.  Discussion Paper Nr. 2009-45 (2009)

\bibitem{CP08}
A.~Chakraborti, M. Patriarca. {Gamma--distribution and wealth inequality}, {\em Pramana J. Phys. } \textbf{71} (2), 233--243 (2008).

\bibitem{ChChSt05} A.~Chatterjee, B.K.~Chakrabarti,
  R.B.~Stinchcombe. Master equation for a kinetic model of trading
  market and its analytic solution. {\em Phys. Rev. E} \textbf{72}, 026126 (2005).


\bibitem{CPP}
S. Cordier, L. Pareschi and C. Piatecki. Mesoscopic modelling of
financial markets. \emph{J. Stat. Phys.} \textbf{134} (1), 161--184
(2009)

\bibitem{CoPaTo05} S.~Cordier, L.~Pareschi, G.~Toscani. On a kinetic
  model for a simple market economy. {\em J. Stat. Phys.} \textbf{120}, 253-277 (2005).

\bibitem{DY00}
A.~Dr\v{a}gulescu, V.M.~Yakovenko, {Statistical mechanics of money},
{\em Eur. Phys. Jour. B} \textbf{17}, 723-729 (2000).


\bibitem{DMT}
B.~D{\"u}ring, D.~Matthes, G.~Toscani. Kinetic Equations modelling Wealth Redistribution:
A comparison of Approaches, \emph{Phys. Rev. E},
\textbf{78}, (2008) 056103

\bibitem{DMT1}
B.~D{\"u}ring, D.~Matthes, G.~Toscani. A Boltzmann-type approach to
the formation of wealth distribution curves,  (Notes of the Porto
Ercole School, June 2008)   \emph{Riv. Mat. Univ. Parma} (1)
\textbf{8}  199-261 (2009).


\bibitem{Ligg}
R. Durrett, T. Liggett
 Fixed points of the smoothing transformation.
 \textit{Z. Wahrsch. Verw. Gebiete}  {\bf 64}  275--301 (1983).


\bibitem{fristedgray}
B.  Fristedt and L.  Gray.
  \textit{A Modern Approach to Probability Theory}.
  Birkh\"auser, Boston, (1997).



\bibitem{GSV} U. Garibaldi, E. Scalas and P. Viarengo. Statistical equilibrium
in simple exchange games II. The redistribution game. \emph{Eur.
Phys. Jour. B}  \textbf{60}(2)  241--246 (2007).



\bibitem{gradshteyn}
 I.~S. Gradshteyn,  I.~M. Ryzhik. \textit{Table of integrals, series,
  and products}, San Diego, Academic Press Inc., 6th ed. (2000)



\bibitem{gup}
A.K. Gupta Models of wealth distributions: a perspective. In \emph{Econophysics and sociophysics: trends and perspectives}
 B.K. Chakrabarti, A. Chakraborti, A. Chatterjee (Eds.) Wiley VHC, Weinheim 161-190 2006.

\bibitem{Gua}
S. Guala. Taxes in a simple wealth distribution model by inelastically scattering particles, \emph{Interdisciplinary Description of Complex
Systems} \textbf{7}(1), 1-7,  2009.

\bibitem{Ha02}
B.~Hayes, {Follow the money}, {\em American Scientist} \textbf{90} (5), 400-405 (2002).

\bibitem{IKR98}
S.~Ispolatov, P.L.~Krapivsky, S.~Redner, {Wealth distributions in
  asset exchange models}, {\em Eur. Phys. Jour. B} \textbf{2}, 267-276 (1998).

\bibitem{Liu}
 Q. Liu.
 On generalized multiplicative cascades.
 \textit{Stochastic Process. Appl.} {\bf 86}  263--286 (2000).


\bibitem{MaTo07} D.~Matthes, G.~Toscani. On steady
distributions of kinetic models of conservative economies. {\em
  J. Stat. Phys.} \textbf{130}, 1087-1117 (2008).

\bibitem{Pat}
M. Patriarca, E. Heinsalu,  A. Chakraborti. Basic kinetic wealth-exchange models: common features and open problems. \emph{Eur. Phys. J. B}
\textbf{73}, 145–153 (2010).

\bibitem{Pia}
S. Pianegonda, J.R. Iglesias, G. Abramson, J.L. Vega. Wealth redistribution with
finite resources. \emph{Physica A} \textbf{322} (2003) 667--675

\bibitem{Sl04}
F.~Slanina, {Inelastically scattering particles and wealth distribution in an
open economy}, {\em Phys. Rev. E} \textbf{69}, 046102 (2004).

\bibitem{steutel} F. W. Steutel and K. van Harn.
{\em Infinite divisibility of probability distributions on the real line}. Marcel Dekker, Inc., New York, (2004)

\bibitem{To2}
G. Toscani, Wealth redistribution in conservative linear kinetic models with taxation, \emph{Europhysics Letters} \textbf{88} (1) (2009) 10007

\bibitem{yak}
V.M. Yakovenko, Econophysics, Statistical Mechanics Approach to, \emph{Encyclopedia of Complexity and System Science}, Springer
http://refworks.springer. (2007)

\bibitem{YR}
V.M. Yakovenko, J.B. Rosser. Colloquium: Statistical mechanics of money, wealth and income. \emph{Rev. Mod. Phys.} (to be published); also
available at arXiv: 0905.1518 (2009).

\end{thebibliography}

\end{document}